\documentclass{ametsocV6.1}

\usepackage{gensymb}
\usepackage{graphicx}

\usepackage{caption}

\title{On the predictability of springtime ozone depletion events using the ECCC Global Deterministic Prediction System}

\authors{Jean  de Grandpr\'e \aff{1},\correspondingauthor{Jean de Grandpr\'e, jean.degrandpre@ec.gc.ca}
Irena Ivanova \aff{1},
Yves J. Rochon \aff{2},
Caroline Jouan \aff{3} \thanks{Author Three now at the CICERO Center for International Climate Research, Oslo, Norway}
and~Paul~A.~Vaillancourt~\aff{4}
}

\affiliation{\aff{1}{Air Quality Research Division, Environment and Climate Change Canada, Dorval, Canada} \aff{2}{Air Quality Research Division, Environment and Climate Change Canada, Downsview, Canada} \aff{3}{Canadian Meteorological Center, Environment and Climate Change Canada, Dorval, Canada} \aff{4}{Meteorological Research Division, Environment and Climate Change Canada, Dorval, Canada}
}

\abstract {Ozone depletion events are recurring phenomena in both
  polar regions, characterized by significant interannual variability.
  In this study, the Environment and Climate Change Canada (ECCC)
  Global Deterministic Prediction System is used to investigate the
  medium-range predictability of ozone and weather throughout the
  anomalous polar ozone depletion events of 2020. The system includes
  ozone assimilation and makes use of a prognostic ozone field for the
  computation of heating rates. The ozone scheme uses simplified
  photochemical modules to represent the impact of both gas-phase and
  heterogeneous reactions throughout polar ozone depletion events.
  The study shows that during the Boreal and Austral spring seasons,
  the predictability of the total ozone column exceeds 10 days and is
  comparable to the predictability of large-scale weather
  variables. It also demonstrates that over both polar regions, the
  inclusion of ozone radiative coupling has a significant impact on
  the temperature and wind distributions throughout the
  stratosphere. Over Antarctica, the ozone coupled forecasts are
  systematically colder at all lead times, which helps eliminate a
  temperature bias present in the model using climatological
  ozone. The strength of the polar vortex also increases significantly
  throughout the lower stratosphere, in better agreement with zonal
  wind analyses. Over the Arctic the use of an ozone-interactive model
  also produces significant changes in the temperature and wind
  forecasts, but the impact on the quality of the weather forecasts is
  generally neutral. The study shows the overall benefits of using
  ozone coupled models in the highly perturbed springtime conditions
  of the polar regions.}

\begin{document}

\maketitle


\statement{Ozone hole events which occur during springtime over polar
  regions can be predicted several days ahead using Numerical Weather
  Prediction (NWP) systems. The ozone decrease associated with such
  events reduces the absorption of solar radiation, impacting the
  temperature and wind fields in the lower stratosphere at an altitude
  of about 15-20 km. This study shows that including this process
  within an NWP model can improve weather forecasts in the lower
  stratosphere. The impact of this process has been evaluated during
  the severe springtime ozone depletion events that took place in both
  polar regions in 2020. The study demonstrates that over Antarctica,
  using an ozone coupled model produces a stronger and colder
  stratospheric polar vortex in better agreement with temperature and
  wind analyses. Overall, the study highlights the benefits of using
  ozone coupled models in the highly perturbed springtime conditions
  of the polar regions.}

\section{Introduction} 

Even though the ozone layer is slowly recovering after several decades
of decline \citep{WMO2018}, it remains fragile to extreme
meteorological conditions. Indeed, the 2020 springtime ozone depletion
events in both hemispheres have been among the strongest on record in
terms of depth, size, and duration
\citep{Lawrence2020,Manney2020,Inness2020}. Forecasting such events is
important for populations living at high latitudes, who are
potentially exposed to harmful UV radiation
\citep{Hand2016,Witze2020}. Strengthening our forecasting capabilities
to address such issues will also contribute to improving our
comprehension of these phenomena.

The formation of ozone holes is driven by catalytic reactions
involving short-lived halogen compounds. The presence of these species
within the stratospheric polar regions is due to the photodissociation
of chlorofluorocarbons and halons released from anthropogenic
activities. The conversion of these long-lived constituents into
chemically active species takes place through heterogeneous reactions
occurring on the surface of Polar Stratospheric Clouds (PSC)
\citep{Solomon1999}. The ozone loss associated with these processes
can be resolved using different types of models with various levels of
complexity \citep{geer2006,geer2007}.  In the context of NWP
applications, they can be represented with simplified approaches as a
way to minimize the usage of computing resources
\citep{Eskes2002,Sekiyama2005,Stajner2006,flemming2011,Monge2011}.

The predictability of ozone varies significantly during the various
stages of ozone depletion events \citep{flemming2011}. The quality of
ozone forecasts at synoptic timescales depends primarily on the
availability of comprehensive analyses that serve as initial
conditions. During the deepening of the ozone hole, it also depends on
the representation of photochemical processes, whereas the
representation of ozone transport is particularly important during the
ozone recovery period. The predictability of ozone throughout these
events also relies on the quality of weather forecasts, which
determine the temperature and wind distributions.

Ozone forecasting systems are generally based on comprehensive NWP
models that include a lid well above the stratopause and incorporate
different parameterization schemes for representing radiative and
gravity wave drag processes. These components are necessary for
resolving the Brewer-Dobson Circulation, which determines the seasonal
and interannual variability of the middle atmosphere
\citep{Scaife2022}. The use of high-top models is valuable for
improving the representation of other phenomena such as Sudden
Stratospheric Warmings \citep{Baldwin2021}, Arctic Oscillations, and
Quasi-Biennial Oscillations \citep{Scaife2014,Wang2022}, which also
contribute to increasing the weather and ozone predictability in the
stratosphere and below through downward control
\citep{Haynes1991,Baldwin1999}.

The use of ozone as a prognostic constituent in NWP systems allows us
to resolve the complex dynamical, radiative, and photochemical
interactions that take place during ozone hole phenomena
\citep{Haase2019,menard2019,Monge2022,degrandpre2009}. The radiative
response is directly associated with the decrease in solar heating
resulting from the rapid disappearance of ozone within the polar
vortex. The associated cooling contributes to the enhancement of the
ozone depletion process, which enhances the Polar Night Jet (PNJ)
through thermal wind balance. This mechanism decreases the mixing of
low ozone within the polar vortex with ozone-rich air from the lower
latitudes, which in turn contributes to further cooling of the polar
vortex. This positive feedback mainly occurs under the conditions of a
strong PNJ, often prevailing during Southern Hemisphere springtime
periods. Climate studies have shown that resolving the non-zonal
component of the radiative forcing generally reduces ozone heating and
produces a significant delay in the final breakup of the vortex
\citep{Crook2008,Gillet2009}.

The strengthening of the polar vortex can also generate a dynamical
response through its impact on the propagation of planetary waves in
the middle atmosphere \citep{Andrews1987,Charney1961}. The nature of
the response will depend on the strength of the PNJ, varying from year
to year in the polar regions. In the Southern Hemisphere, the increase
of the PNJ will generally preclude the propagation of large-scale
planetary waves, which can strengthen the initial ozone-induced
radiative forcing.  With the presence of a weak polar vortex, as
frequently occurs in the Arctic, the strengthening of the PNJ can
enhance wave activity and increase the number of wave-breaking events,
contributing to the strengthening of the downwelling branch of the
Brewer-Dobson Circulation.  The associated adiabatic warming in this
case can counteract the original radiative forcing and lead to an
early breakup of the polar vortex \citep{Lin2017}. The nature of these
interactions depends on the preconditioning of the polar vortex
\citep{Lawrence2020b}, which can vary throughout the springtime
period.

The quality of ozone and weather forecasts can be evaluated against
observations and analyses in different regions at different lead times
using standard measures such as Anomaly Correlation Coefficient (ACC)
and root mean square errors \citep{Geer2016}. ACC provides a measure
of predictability in terms of timescales and can be applied to
different fields, including Total Ozone Column (TOC), which serves as
an excellent proxy for characterizing the ozone distribution in the
lower stratosphere. TOC can be measured accurately from satellites and
ground-based instruments, and its predictability is comparable to the
predictability of large-scale weather variables such as temperature
and geopotential height throughout the lower stratosphere. Being an
integrated quantity, TOC needs to be evaluated in combination with
other measures such as volume mixing ratio to assess species vertical
distribution. TOC is widely used for assessing ozone forecasting
systems in the lower stratosphere, where in-situ observations are
sparse \citep{Eskes2002,Sekiyama2005,Stajner2006}.

In this study, the ECCC Global Deterministic Prediction System (GDPS)
is used to determine the ozone and weather predictability during the
Boreal and Austral ozone hole events of 2020. The impact of including
ozone radiative coupling with a simplified heterogeneous chemistry
scheme is evaluated during the different stages of these events. The
system is run in different configurations to isolate the impact of
these different model components on the quality of the weather
forecasts. The GDPS is evaluated in both polar regions at the time
when ozone depletion was particularly intense. The ozone and weather
forecasts of several variables, including temperature and zonal wind,
are evaluated against analyses at different lead times. The study
includes a description of the model and the assimilation system in
Sections 2 and 3, respectively. The results and analyses are presented
in Section 4, and the main conclusions are summarized in Section 5.

\section{The ozone coupled NWP model} \label{sec:model_gem}

The ozone coupled GDPS has been in operation at ECCC since November
2021, making it the first global NWP system to incorporate ozone
radiative coupling for operational use.  This system is based on the
Global Environmental Multiscale (GEM) NWP model, which has a
dynamical core that solves the fully compressible governing equations
using an implicit two-time level Semi-Lagrangian advection scheme
\citep{Cote1998,Girard2014,Husain2017}. The current operational system
runs at 15 km resolution on 84 levels with a lid at 0.1 hPa
\citep{charron2012}. All model configurations employ the Arakawa-C
staggering on a global latitude-longitude grid using the Yin-Yang
strategy, represented by a pair of overlapping limited-area grids
\citep{QaddouriLee2011}. Vertically, equations are discretized on a
terrain-following log-hydrostatic-pressure coordinate
\citep{Girard2014}.

In this study, all experiments are conducted with the same model
configuration as the one utilized in operation, except for the use of
a 25 km horizontal resolution to save computing resources.  The
physical parameterization packages are from \citet{McTaggart2019},
which are used by the global and regional operational weather
forecasting systems.  It includes a non-orographic gravity wave drag
parameterization scheme \citep{Hines1997} that incorporates sources
initiated by imbalances and geostrophic adjustments related to
convection, fronts, and other transient mesoscale phenomena. Radiative
processes are parameterized with the correlated-k distribution
approach following \citet{LiBarker2005}, which solves radiative
transfer equations in nine longwave and four shortwave spectral
bands. For the computation of heating rates, the model uses a
prognostic ozone representation based on a simplified photochemical
scheme described below.

\subsection{Linearized stratospheric ozone chemistry and equilibrium state}

The use of linearized schemes in NWP models allows for the inclusion
of a prognostic representation of ozone while minimizing the CPU load
on supercomputers \citep{geer2007}. These approaches use comprehensive
models to generate a set of coefficients, which are then used as input
to a linearized model. Several linearized schemes are available for
including ozone as a prognostic constituent in NWP systems, such as CD
\citep{Cariolle2007}, CHEM2D \citep{McCormack2006}, LINOZ
\citep{McLinden2000}, and BMS \citep{Monge2011}. These schemes vary in
complexity and are used to linearize various processes that determine
the ozone distribution throughout the model domain. The CD and CHEM2D
schemes use 2-D models for linearizing species transport and gas-phase
photochemical processes throughout the stratospheric domain. The LINOZ
scheme uses a comprehensive box model that only serves for the
linearization of gas-phase reactions.

The CD, CHEM2D, and LINOZ schemes do not incorporate heterogeneous
reactions into the linearization process. The impact of these
reactions needs to be parameterized as an additional process for the
study of polar ozone depletion issues. The simplified parameterization
schemes used to address such issues generally use prescribed
parameters to represent the chemical ozone loss within the polar
vortex. They may require some tuning procedures to adjust the ozone
loss rates according to a specific timescale within a restricted
area. As an alternative, the BMS scheme uses a more comprehensive
chemical transport model that includes both gas-phase and
heterogeneous reactions for the generation of linear
coefficients. Including heterogeneous reactions in the linearization
process avoids the need to develop additional photochemical
parameterization schemes. The BMS approach can be used for a wide
range of applications \citep{Monge2022}, making it a valuable
alternative in the context of operational NWP forecasting.

At NWP timescales, modeling errors generally increase with time and
depend on species’ initial conditions. In this study, the length of
the forecasts is limited to 10 days, and ozone analyses from the GDPS
are used as initial conditions. This experimental framework prevents
the development of large forecast errors and makes it possible to use
a simplified ozone modeling approach for investigating polar ozone
depletion processes. The prognostic ozone scheme used in this study is
based on the LINOZ approach, which runs in-line within the GEM NWP
model. Following this approach, species transport is resolved
explicitly by the model using a Semi-Lagrangian advection scheme
\citep{degrandpre2016}. Ozone tendencies are represented by the
first-order Taylor expansion about stratospheric chemical rates and
depend on the ozone mixing ratio ($\phi$), the temperature (T), and the
overhead column ozone c, expressed as follows:

\begin{equation}
\frac{d \phi}{dt} =
(P-L)^{o} +
\frac{\partial (P-L)}{\partial \phi}|_{o} (\phi-\phi^{o}) +
\frac{\partial (P-L)}{\partial T}|_{o} (T -T^{o}) +
\frac{\partial (P-L)}{\partial c}|_{o} (c-c^{o}),
\label{eq:o3}
\end{equation}

\noindent where superscripts $^{o}$ represent climatological values
and $|_{o}$ denotes the evaluation of partial derivatives with respect
to the climatological quantities $\phi^{o}$, $T^{o}$ and $c^{o}$.  The
gas-phase photochemical production and loss terms (P and L) have been
computed offline using a comprehensive box model as described in
\citet{McLinden2000}. All other coefficients have been updated
following \citet{Hsu2009}. Below the tropopause, defined as the level
where water vapor becomes greater than 10 ppmv, ozone is advected as a
passive tracer. Above 1 hPa and below 400 hPa, the ozone mixing ratio
is forced toward the model climatology $\phi^{o}$ following the
expression:

\begin{equation}
\phi (t+\Delta t) =
(1 - \frac {\Delta t}{\tau_{chem}}) \phi(t^{*}) +
(\frac{\Delta t}{\tau_{chem}}) \phi^{o} .
\label{eq:o3b}
\end{equation}

\noindent In this equation $\phi(t^{*})$ represents species mixing
ratio at the intermediate time level $t^{*}$ following the application
of the photochemical sink/sources.  $\tau_{chem}$ is a chemical
relaxation timescale here chosen as 2 days in the troposphere and 6
hours above 1 hPa. $\Delta t$ is the model timestep.

The choice of ozone and temperature climatologies in equation
(\ref{eq:o3}) can affect the quality of the ozone and weather
forecasts in several ways. These climatologies are used for generating
linear coefficients and for constraining the ozone distribution
outside the stratospheric and UTLS domains. They also determine the
ozone photochemical response to temperature and ozone perturbations,
which is a fundamental issue with the use of ozone coupled models. The
ozone prognostic equation in (\ref{eq:o3}) shows that the
photochemical term is particularly sensitive to the choice of
climatologies. It can alter the interactions between ozone and
temperature via unphysical contributions from the perturbation terms
on the RHS of the equation. To solve these radiative and photochemical
interactions, it is necessary to choose a set of ozone and temperature
climatologies that represent a state of equilibrium representative of
stratospheric conditions.

Climatologies can be built from different sources, including
observations, reference models, and reanalyses. In this study, the
original observation-based climatologies used by the LINOZ scheme have
been replaced by a new set of climatologies generated from multi-year
reanalyses. The use of reanalyses helps preserve the photochemical and
radiative balance between ozone and temperature climatologies, which
cannot generally be represented using observation-based datasets. With
reanalyses, the ozone and temperature climatologies can be generated
from the same dataset over the same domain and period, increasing the
level of consistency between these two fields. The use of reanalyses
also allows for updating the ozone distribution by taking into account
changes that have occurred over the past decades \citep{WMO2018}.

Different types of reanalyses, such as ERA5 \citep{Hersbach2020},
M2-SCREAM \citep{Wargan2023}, and CAMS \citep{Inness2019}, can be used
to constrain the ozone linearized scheme throughout the stratospheric
and tropospheric domains. The M2-SCREAM and CAMS assimilation systems
provide a comprehensive set of chemical analyses that can be used to
constrain the ozone distribution in various domains. The ERA5 system
is based on a linearized ozone scheme but includes ozone as a model
variable within the 4Dvar assimilation system. This approach
contributes to preserving the coherence between the ozone and
temperature analyses, which is an important element of the study.

In this study, we have chosen ERA5 reanalyses to maximize the
consistency between ozone and temperature climatologies. The monthly
mean ozone and temperature have been generated from the same datasets
for the period 2010 to 2016. This period is relatively short but
comparable to the period covered by other climatologies used in
linearized schemes \citep{geer2007}. In the lower mesosphere, from 1
hPa to 0.3 hPa, the ozone climatology is based on the
\citet{Fortuin1998} reference model, whereas ozone measurements from
the Halogen Occultation Experiment (HALOE) are used from 0.3 hPa to
the model lid. The ozone climatology is interpolated on hybrid model
levels via linear interpolation in log pressure coordinates.

\subsection{Heterogeneous processes}

The model includes a parameterization scheme for representing the
photochemical reactions occurring on the surface of PSCs. The
formulation is based on a cold tracer approach for determining the
pre-conditioning phase of air masses (e.g. \citet{Cariolle2007} and
\citet{geer2007}). According to these approaches, the degree of
halogen activation is determined by a tracer $\beta$ , which is used
to parameterize the rate of ozone destruction, taking the following
form:

\begin{equation}
\frac{d \phi}{dt}|_{het} = \frac{-1}{\eta} \beta \phi .
\label{eq:hetchem}
\end{equation}

\noindent This expression is only applied in direct sunlight conditions using
$\eta$ = 10 days as a decaying time constant representative of the
strength of the ozone depletion process.  The prognostic equation for
the cold tracer $\beta$ is defined as:

\begin{equation}
\frac{d \beta}{dt} = \frac{1}{\eta_{P}} (1 -\beta) - \frac{1}{\eta_{L}} \beta
\label{eq:beta}
\end{equation}

\noindent following \citet{geer2007}.  The first term on the right
hand side represents the activation process and is applied when the
temperature is below a threshold value of 195K. $\eta_{P}$ represents
a characteristic activation timescale estimated as 4 hours. The second
term represents the de-activation process and is applied in direct
sunlight conditions with a time constant $\eta_{L}$ of 10 days.
$\beta$ is reinitialized to nil at the beginning of each forecasts.

The ozone tendency in (\ref{eq:hetchem}) is added after applying the
gas-phase tendency in (\ref{eq:o3}) using a process splitting
approach. The parameterization scheme is applied in a restricted area
between 12 km and 22 km based on a sensitivity study that indicated a
slight overestimation of the ozone-depleted area when the original
settings from \citet{geer2007} were used.  This specific choice of
settings appears appropriate in the context of the very severe 2020
ozone depletion events. However, further evaluation is needed to
ensure that the parameterization scheme can also represent the large
interannual variability of ozone loss processes in both polar regions
\citep{Monge2022}.


\section{Ozone assimilation}

The GDPS assimilates both meteorological and ozone
observations. Meteorological observations are assimilated using the
ensemble–variational approach \citep{Buehner2015}, whereas a
three-dimensional variational assimilation with a first-guess at the
appropriate time is used for ozone assimilation \citep{Rochon2019}.
This difference arises because ozone forecasting is not included in
the system that generates the ensembles necessary for the
ensemble-variational approach.

In this study, ozone retrievals from six satellite instruments are
assimilated. The bulk of the data consists of retrieved integrated TOC
values from the Ozone Monitoring Instrument (OMI) on board the AURA
satellite \citep{Bhartia2002}, the nadir mapping from the Ozone
Mapping and Profiler Suite (OMPS-NM) on board the Suomi National
Polar-orbiting Partnership (Suomi NPP) \citep{Flynn2017}, and
measurements from the Tropospheric Monitoring Instrument (TROPOMI) on
board Sentinel-5 \citep{Garane2019}. The assigned observation error
standard deviation for OMI and OMPS-NM data is set from a single
iteration of the \citet{Desroziers2005} approach, with verification
and adjustment from the standard deviation of differences from
colocated OMI and OMPS-NM data. Those for TROPOMI are as provided with
the level 2 data.

Ozone data from three profiler instruments are also assimilated,
including measurements from the Microwave Limb Sounder (MLS) on board
AURA, measurements from the Backscatter Ultraviolet second-generation
Instrument (SBUV/2) on board NOAA-19, and measurements from the Nadir
Profiler of the Ozone Mapping and Profiler Suite (OMPS-NP) on board
Suomi NPP. MLS retrieved concentration profiles are most effective in
constraining and preserving the quality of the ozone field’s vertical
structure. It provides data under both daytime and nighttime
conditions, while all others only provide daytime data. The influence
of the SBUV/2 and OMPS-NP retrieved partial column profilers is
currently weak, even more so for the latter which remains to be
resolved. References on retrievals for the profiler data are
summarized in \citet{Peters2013} for SBUV/2 and \citet{Bai2016} for
OMPS-NP. The applied observation error standard deviation of the
profiles is currently those provided with the retrieval products.

The 21 layers of the original partial column profiles are reduced to
six thicker, potentially usable layers, as these profiles contain the
equivalent of at most about four pieces of independent
information. The corresponding retrieved measurement averaging kernels
are applied in the assimilation. The roughly four pieces of
independent information denote the reduced degree of freedom resulting
from the correlation of the original layers, as seen through the
overlapping averaging kernels of the retrieved measurements. The
thicker layers are generated by combining neighboring original
layers. The result is a set of layers that are less correlated with
each other through their averaging kernels, except for the strongly
coupled bottom two layers.  The specified layer boundaries are 0.101,
1.01, 2.54, 6.38, 25.43, 254.3 hPa, and the surface.

\section{Chemical data assimilation experiments}

The ozone coupled GDPS was evaluated during the 2020 Northern
Hemisphere (NH) and Southern Hemisphere (SH) ozone hole events. The
system produced weather and ozone analyses every 6 hours, while 10-day
forecasts were launched every 12 hours at a 25 km resolution
throughout both springtime periods. The assimilation cycle was based
on a model version that included ozone radiative coupling but did not
include the parameterization of heterogeneous chemistry. Past
experiments with the GDPS system have shown that the impact of the
parameterization scheme on the quality of ozone analyses is not
significant with the use of a 6-hour assimilation window. This is
attributed to the fact that the number of observations used by the
assimilation system is generally large enough to produce a realistic
representation of the ozone hole throughout the entire springtime
period.

The weather and ozone analyses produced by the GDPS system serve as
initial conditions for all forecast experiments. They also serve to
evaluate the quality of the forecasts at different lead times. All
experiments consist of launching a single 10-day forecast every 12
hours throughout both springtime periods. In this study, forecast
experiments with and without the heterogeneous parameterization scheme
have been performed to evaluate the impact of those reactions on ozone
predictability. Further experiments with and without ozone radiative
coupling have been performed to evaluate the radiative impact of ozone
on the quality of the weather forecasts.

\subsection{Ozone forecasting}

\subsubsection{Southern Hemisphere}

In 2020, the ozone hole over Antarctica was one of the strongest on
record in terms of size and depth (Blunden, 2021). It was also one of
the longest lasting events, with the breaking of the vortex occurring
at the end of December. During this period, the minimum Total Ozone
Column (TOC) was diagnosed daily to evaluate the capability of the
heterogeneous chemistry parameterization to deplete ozone below the
220 DU threshold, which is considered the minimum value that can be
obtained from dynamical processes.

\begin{figure}
\includegraphics[width=16cm]{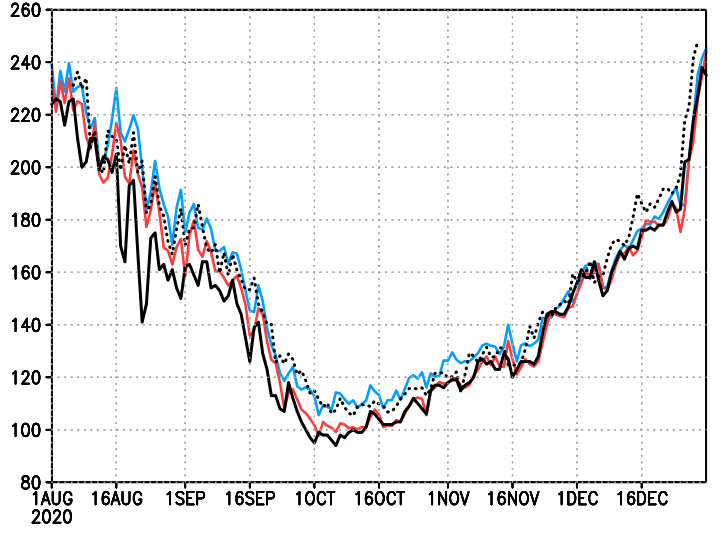}
\caption{Time Series of the TOC minimum (DU) from August 1$^{st}$ to
  December 31$^{th}$ 2020 within the 40\degree S and 90\degree S
  latitude bands. From GDPS analyses (red), CAMS reanalyses (blue),
  observations from NASA Ozone Watch (black) and GDPS forecast at
  5-day lead time (black dotted line)}
\label{colo3_sp}
\end{figure}

Figure~\ref{colo3_sp} shows the evolution of the TOC minimum between
the 40\degree S and 90\degree S latitude bands from August 1$^{st}$ to
the end of December 2020.  Results are diagnosed from the GDPS
analyses, the Copernicus Atmosphere Monitoring Service (CAMS)
reanalyses \citep{Inness2019}, and measurements from the OMPS
instrument onboard the Suomi NPP satellite, taken from the NASA Ozone
Watch website at https://ozonewatch.gsfc.nasa.gov. The minimum value
is picked within a wide latitude band for all the datasets to ensure
that the diagnostic is applied throughout the whole area of the polar
vortex for the entire event.  Data reveal that both analysis systems
effectively capture the ozone behavior from the ozone-hole formation
period to the recovery period beginning in early October.  From August
to mid-September, the OMPS values are significantly lower than GDPS
and CAMS, which could be partially attributed to the limited amount of
ozone observations that can be assimilated over the period in the
absence of sunlight. Following this period, there is good agreement
between the GDPS analyses and the OMPS instrument, whereas the ozone
minimum from CAMS is slightly higher.

The differences observed in the timeseries between the GDPS and CAMS
can be attributed to various assimilation and modeling aspects. The
CAMS reanalysis system is based on a model that includes the
parameterization of heterogeneous reactions \citep{Cariolle2007},
whereas the GDPS assimilation cycle is based on a model version that
does not include such processes. Nevertheless, the GDPS system
produces lower ozone values, highlighting that the differences
observed throughout the period are likely associated with the ozone
assimilation process. The use of a bias correction scheme in the CAMS
system, but not in the GDPS, is one factor that could produce
systematic differences between both assimilation systems.

Figure~\ref{colo3_sp} also presents the results from the forecast
experiment at a 5-day lead time using a model version that includes
the parameterization of heterogeneous reactions. It indicates that the
differences between the ozone forecasts and analyses generally remain
under 10 DU, which is within the analysis uncertainties estimated from
the differences between the GDPS and CAMS analyses.

Throughout August and September, the ozone forecasts are in
general agreement with analyses, indicating that the parameterization
scheme provides a reasonable estimation of the ozone loss
process. During October, the ozone forecast systematically
overestimates the ozone minimum, which can be attributed to numerical
effects such as model resolution, which may not be sufficient to
preserve the TOC minimum values present in the analyses. In November
and December, the ozone layer recovers rapidly due to the mixing of
ozone-depleted air within the polar vortex with ozone-rich air from
the mid-latitudes. This large-scale process appears to be reasonably
represented by the model, even though the model overestimates the
ozone recovery associated with the final breakup of the polar vortex
in December.

\begin{figure}
\includegraphics[width=16cm]{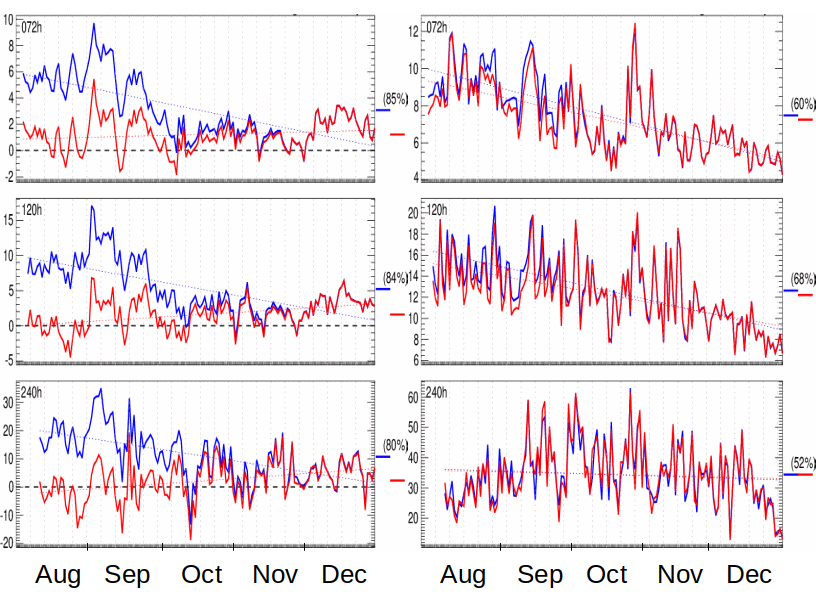}
\caption{Time Series of TOC mean differences (left) and standard
  deviation (right) in (\%) between forecasts and GDPS analyses in the
  60\degree S and 90\degree S latitude bands at 70 hPa.  Ozone
  forecasts at 3-5-10-day lead times are from the experiments with
  (red) and without (blue) the parametrization of heterogeneous
  chemistry.  Marks on the right axis denote the mean differences (\%)
  and the number in brackets denotes the percentage of forecasts for
  which the differences are larger for the experiment without the
  parameterization scheme.  The dotted lines are the linear regression
  that best fits the data of both experiments.}
\label{ts_o3du_sp}
\end{figure}

In Figure~\ref{ts_o3du_sp}, the experiment is compared against a
control in which the ozone forecasts were launched without including
the parameterization of heterogeneous reactions. The left panels
present the time evolution of TOC mean bias (\%) evaluated against the
GDPS analyses over the Antarctic region at different forecast lead
times. The figure shows that the experiment without the
parameterization scheme suffers from a significant bias, which is
particularly important during the ozone hole formation period when the
strength of the chemical ozone loss maximizes. Results indicate that
the mean bias throughout the entire period increases linearly from 3\%
at a 3-day lead time to about 10\% at a 10-day lead time. With the
inclusion of heterogeneous chemistry, the mean bias decreases
significantly and remains under 2\% for all lead times. This result
suggests that the choice of ozone depletion lifetime in the
heterogeneous parameterization scheme is slightly more appropriate for
capturing the strength of the loss rate at longer timescales. During
the vortex closure period starting from mid-October, the impact of the
photochemistry is smaller, and both experiments give similar results.

The right panels in Figure~\ref{ts_o3du_sp} present the time evolution
of TOC standard deviation over the region.

The standard deviation appears larger in the early stages of the event
and decreases progressively over time. This trend is particularly
significant at a 5-day lead time, showing a constant decline in the
standard deviation from about 16\% in early August to 8\% at the end
of December. This systematic reduction is attributed to a better
representation of dynamical processes as we progress toward the vortex
breakup period. In August and September, the TOC standard deviation is
slightly improved due to the inclusion of heterogeneous chemistry.
During the ozone recovery period from mid-October, the impact of the
parameterization scheme on the standard deviation is not significant,
and the model errors appear mainly determined by the representation of
dynamical processes, which could be improved with the use of higher
resolution models. Results also reveal that the standard deviation
decreases significantly at all lead times in late December, indicating
that the GDPS can capture the final breakup of the polar vortex. At
the 10-day lead time, the standard deviation is relatively constant
regardless of the processes involved, suggesting that the gain of
predictability obtained from model developments will likely be reduced
at those timescales.

\begin{figure}
\includegraphics[width=16cm]{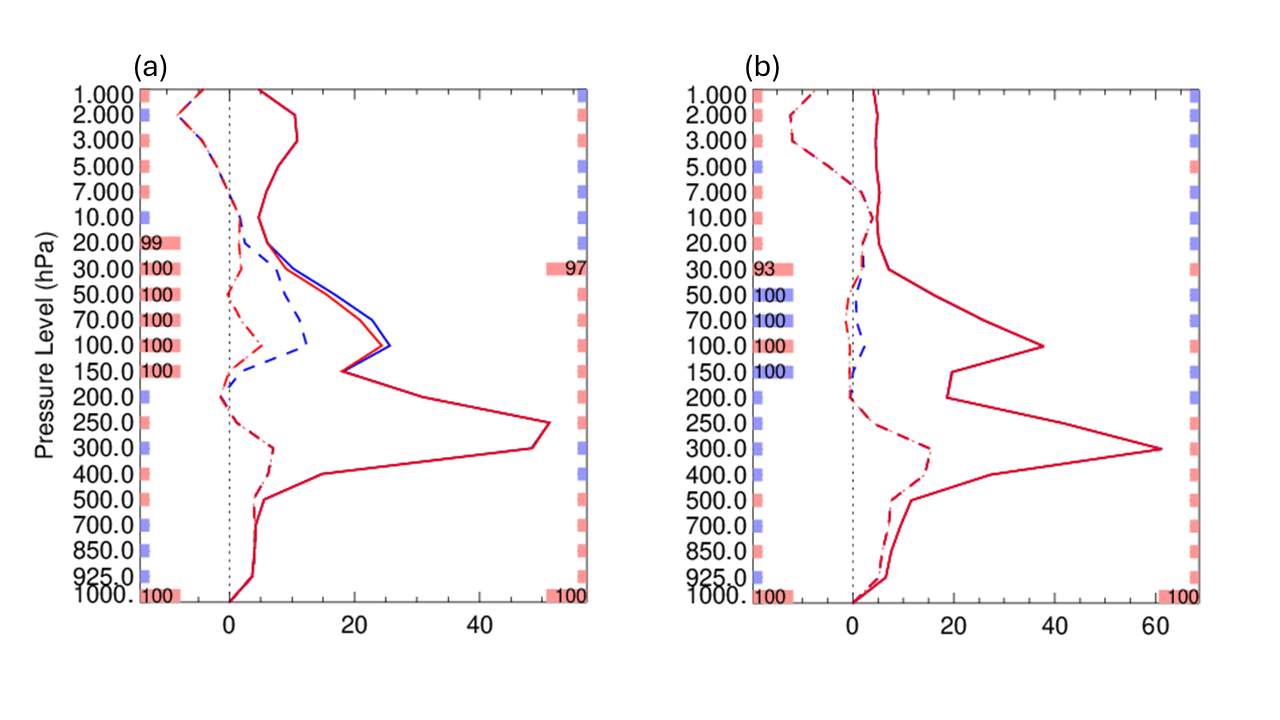}
\caption{Five-day ozone forecasts against GDPS analyses (\%) within
  the 60\degree S and 90\degree S latitude bands from (a) August
  1$^{st}$ to September 30$^{th}$ 2020 and (b) October 1$^{st}$ to
  December 31$^{th}$ 2020.  Ozone forecasts are from the experiments
  with (red) and without (blue) the parametrization of heterogeneous
  chemistry.  Dash lines are the mean and solid lines are the standard
  deviation.  Boxes on the left (right) denote statistical
  significance levels for the bias (standard deviation) computed using
  permutation tests (see chapter 15 of \citet{Efron1994}).  Red (blue)
  boxes mean that the experiment which includes the parameterization
  scheme is better.}
\label{prof_o3_sp}
\end{figure}

The overall impact of the heterogeneous chemistry parameterization
scheme on the ozone forecasts at a 5-day lead time over Antarctica
is presented in Figure~\ref{prof_o3_sp}. From August to September
(left panel), the results show a statistically significant reduction
of the mean ozone bias by about 10\% throughout the lower
stratosphere. The figure also shows a systematic decrease in the ozone
standard deviation below 30 hPa, reaching an average of 2\% in the
lower stratosphere. This reduction does not appear significant at the
95\% confidence level but is observed throughout the lower
stratosphere, where the temperature of the polar vortex is cold enough
for the generation of PSCs. Results show that the ozone standard
deviation is very large in the tropopause region, which may indicate
the limitations associated with the use of a linearized stratospheric
model at those altitudes, as discussed later in section 4a.2. From
October to December (right panel), the ozone bias and standard
deviation increase throughout most of the lower stratosphere, which
can be attributed to the large dynamical variability of the polar
vortex. The signal is statistically significant throughout the region,
but the impact of the parameterization scheme on the quality of the
ozone forecast is generally neutral throughout this period.

\begin{figure}
\includegraphics[width=16cm]{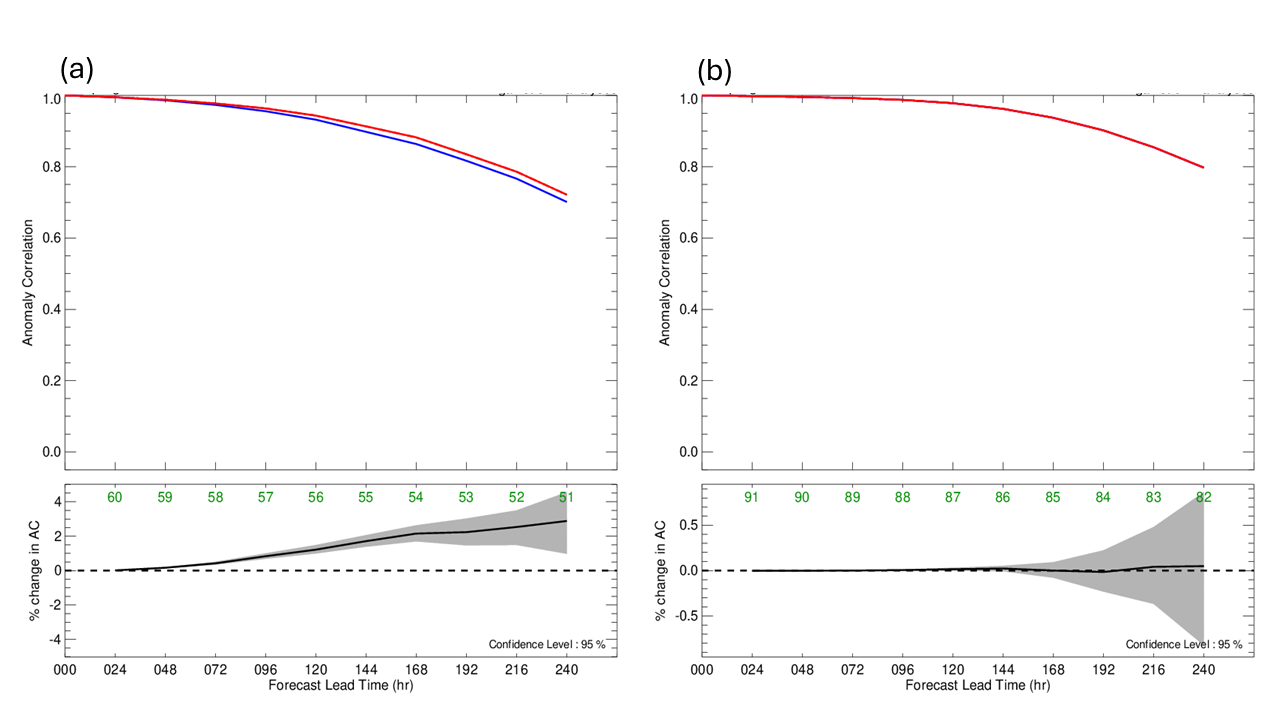}
\caption{Anomaly Correlation Coefficient of TOC for the period (a)
  August 1$^{st}$ to September 30$^{th}$ 2020 and (b) October 1$^{st}$
  to December 31$^{th}$ 2020 as a function of lead times (in hours)
  within the 60\degree S and 90\degree S latitude bands.  Ozone
  forecasts are from the experiments with (red) and without (blue) the
  parametrization of heterogeneous chemistry.  Digits in green
  represent the number of forecasts used in the calculation at the
  given lead time.  The bottom panel shows the differences between
  both experiments whereas the shading indicates the 95\% confidence
  level. Confidence limits are obtained from bootstrapping the
  adjusted correlation coefficient after applying a Fisher
  transformation to convert the correlation coefficient to a normally
  distributed variable. This process involves repeatedly resampling
  the data with replacement.}
\label{ac1_toto3_sp}
\end{figure}

The ozone predictability in the lower stratosphere can be estimated by
comparing TOC forecasts and analyses using ACC. The computation of
ozone anomalies is done at every 24-hour lead time and makes use of
the ERA5 ozone climatology as the reference dataset.
Figure~\ref{ac1_toto3_sp}a presents the TOC predictability during the
ozone hole formation period from August to September when the quality
of the forecast is largely controlled by photochemical processes. The
results show that the inclusion of the parameterization scheme
increases the TOC predictability at all lead times during this
period. After 10 days, the ACC difference reaches 0.04, which
corresponds to a gain of predictability of about 8 hours,
statistically significant at the 95\% confidence level as denoted on
the bottom panel. This highlights the reduction of modeling errors
associated with the use of the parameterization scheme. The ozone
predictability during this period exceeds 10 days using the 0.6 ACC
reference value as a benchmark. This estimation is significantly
larger than other values found in the literature \citep{Sekiyama2005,
  Eskes2002, Stajner2006}, highlighting the progress made in the
performance of operational NWP systems over the past decades.
Figure~\ref{ac1_toto3_sp}b shows that during the vortex closure period
starting in October, the ACC value increases from 0.7 to 0.8 at a
10-day lead time, which corresponds to a gain of predictability of
about 1 day. Throughout this period, the predictability of TOC is
comparable to the predictability of temperature at 70 hPa shown on
Figure~\ref{acc_tt_70hpa_sp}.

\begin{figure}
\includegraphics[width=14cm]{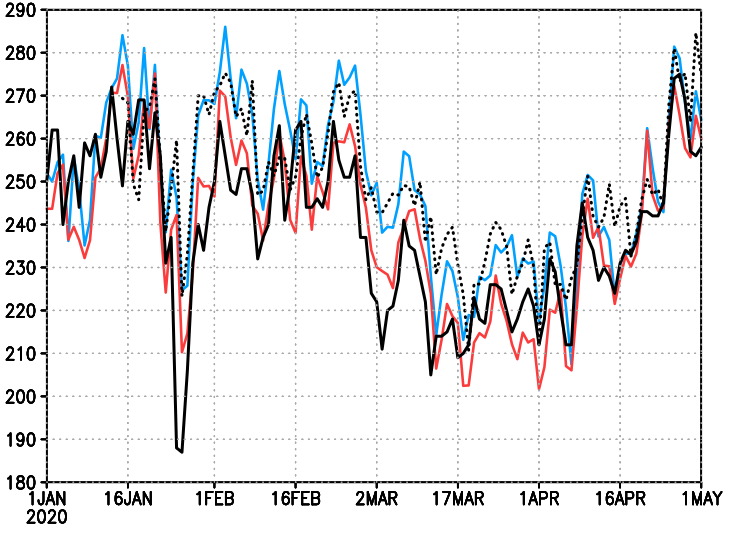}
\caption{Time Series of the TOC minimum (DU) from January 10$^{th}$ to
  April 30 $^{th}$ 2020 within the 60\degree N and 90\degree N
  latitude bands. From GDPS analyses (red), CAMS reanalyses (blue),
  observations from NASA Ozone Watch (black) and GDPS forecast at
  5-day lead time (black dotted line)}
\label{colo3_np}
\end{figure}

\subsubsection{Northern Hemisphere}

The system has been evaluated in the northern polar region during the
late winter and spring periods of 2020, when the temperature stayed
below the PSC threshold value for several weeks. The large
spatio-temporal variability of the region provides suitable conditions
for assessing ozone predictability in a dynamical regime where
chemical loss rates are relatively small compared to the SH.

Figure~\ref{colo3_np} shows the evolution of the TOC minimum between
the 60\degree N and 90\degree N latitude bands from January 10$^{th}$
until the end of April 2020. Results are diagnosed as previously from
the GDPS analyses, the Copernicus Atmosphere Monitoring Service (CAMS)
reanalyses, and measurements from the OMPS instrument onboard the
Suomi NPP satellite, taken from the NASA Ozone Watch website. During
this period, both sets of analyses reproduce the strong day-to-day
variability, with minimum ozone values just above 200 DU towards the
second half of March. The mean minimum value in March from the GDPS,
at approximately 215 DU, is lower than CAMS by almost 10 DU but is in
general agreement with OMPS and measurements from the TROPOMI
instrument onboard the Sentinel 5 satellite during the same period
\citep{Dameris2021}. The forecasts at a 5-day lead time shown in
Figure~\ref{colo3_np} appear slightly biased against the GDPS analyses
but remain within the range of analysis uncertainties.  As discussed
previously on Figure~\ref{colo3_sp} this systematic overestimation of
the ozone minimum could be partially attributed to a model resolution
issue.

\begin{figure}
\includegraphics[width=16cm]{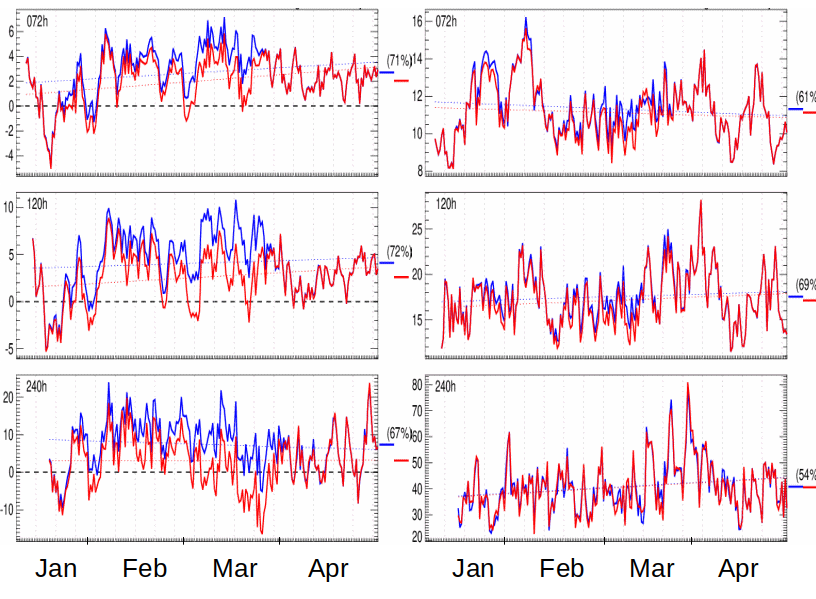}
\caption{Time Series of TOC mean differences (left) and standard
  deviation (right) in (\%) between forecasts and GDPS analyses in the
  60\degree N and 90\degree N latitude bands at 70 hPa. Ozone forecast at
  3-5-10-day lead times from the experiments with (red) and without
  (blue) the parametrization of heterogeneous chemistry.  Marks on the
  right axis denote the mean differences (\%) and the number in
  brackets denotes the percentage of forecasts for which differences
  are larger for the experiment without the parameterization scheme.
  The dotted lines are the linear regression that best fits the data
  of both experiments.}
\label{ts_o3du_np}
\end{figure}

The comparison against the control experiment, which doesn’t include
the parameterization of heterogeneous reactions, is shown in
Figure~\ref{ts_o3du_np}. The results highlight a significant reduction
of the mean biases at all lead times associated with the use of the
parameterization scheme. The impact is particularly important in
March, where the TOC bias at a 5-day lead time is reduced by half with
the inclusion of the parameterization scheme. The mean bias for the
full period is 2\% at a 3-day lead time and 3\% at a 10-day lead time,
indicating that the heterogeneous chemistry scheme produces a better
estimation of the ozone loss rates at longer timescales, as seen in
the SH. Figure~\ref{ts_o3du_np} shows that the TOC standard deviation
is larger than it is over Antarctica, which is associated with the
large dynamical variability of the region. The results also show a few
percent reduction of the standard deviation in March directly
associated with the inclusion of the parameterization scheme.

\begin{figure}
\includegraphics[width=16cm]{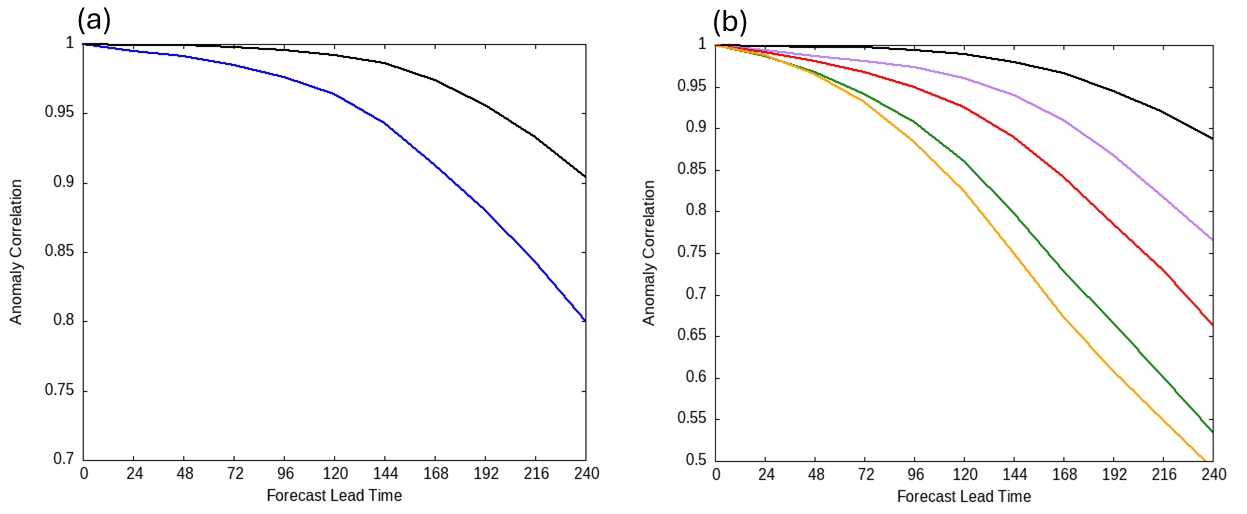}
\caption{Anomaly Correlation Coefficient as a function of lead times
  (in hours) within the 60\degree N and 90\degree N latitude bands
  for the period January 10$^{th}$ to April 29$^{th}$ 2020 from the
  experiment with the parametrization of heterogeneous chemistry. (a)
  Temperature at 70 hpa (black) and TOC (blue). (b) Temperature at 100
  hPa (black) and ozone at 50 hPa (purple), 70 hPa (red), 100 hPa
  (green) and 200 hPa (orange)}
\label{ac2_toto3_np}
\end{figure}

The ozone predictability from January to April remains high and
doesn’t increase significantly when the parameterization of
heterogeneous chemistry is included, as it occurs in the SH. This is
likely associated with the smaller magnitude of the ozone loss process
and the relatively short period in which this photochemical process is
active. Figure~\ref{ac2_toto3_np}a shows that the TOC ACC value
reaches 0.8 after 10 days, which is comparable to the value obtained
during the ozone recovery period in the SH shown previously in
Figure~\ref{ac1_toto3_sp}b. ACC results also indicate that the TOC
predictability is systematically lower than the temperature
predictability at 70 hPa in both hemispheres and decreases faster with
time throughout the lower stratosphere. This is attributed to the fact
that TOC is an integrated quantity, which is strongly influenced by
the ozone distribution in the lowermost stratosphere between 100 hPa
and the tropopause, where most stratospheric-tropospheric exchange
processes take place \citep{Holton1995}.

Figure~\ref{ac2_toto3_np}b shows that ACC values evaluated from the
ozone mixing ratio decrease rapidly below 70 hPa, whereas the
temperature predictability (in black) remains very high throughout the
UTLS region. This rapid decrease in ozone predictability appears at
all latitudes (not shown), highlighting the limitations associated
with the use of a stratosphere linearized scheme in regions
potentially affected by mass fluxes originating from the
troposphere. Results show that the TOC predictability is larger than
the ozone predictability evaluated at specific pressure levels,
indicating that TOC remains a valuable metric for evaluating the
overall performance of the ozone forecasting system. The ozone
predictability below 70 hPa could be improved using higher resolution
forecasts and a more comprehensive photochemical scheme. The
assimilation of additional limb sounding measurements (e.g.
\citet{Flynn2006}) would also contribute to improving the quality of
ozone analyses throughout the region.


\subsection{Ozone radiative coupling}

Throughout springtime periods, the impact of ozone heating on the
evolution of the polar vortices can be significant in both hemispheres
\citep{Monge2022, Oh2022}. In such conditions, the standard approach of
using monthly mean ozone distributions for the computation of heating
rates can generate a systematic overestimation of solar heating, as
demonstrated in \citet{Gillet2009} and \citet{Crook2008}. Furthermore,
this temperature bias can increase throughout the springtime period as
solar forcing becomes more important and the polar vortex becomes more
perturbed and drifts further away from zonal mean conditions.

\begin{figure}
\includegraphics[width=16cm]{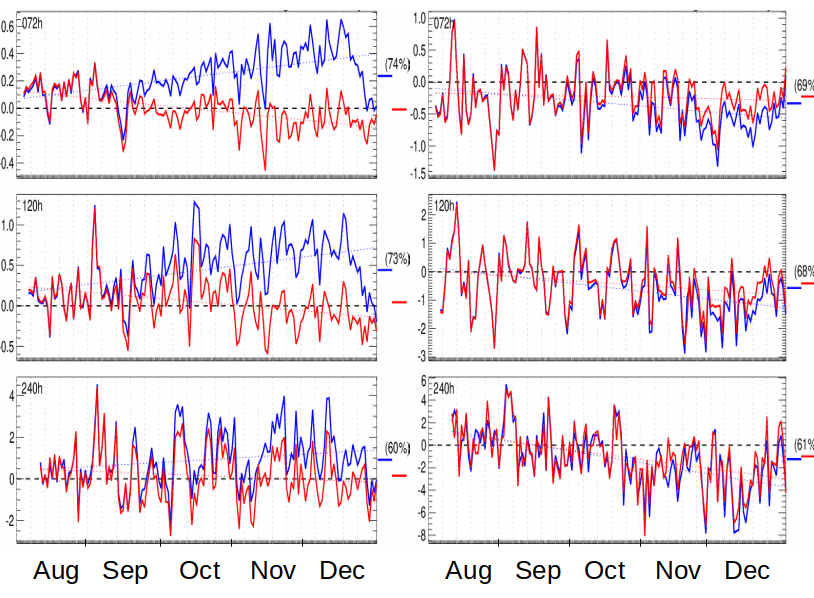}
\caption{Time Series of temperature (left) and zonal wind (right) mean
  differences between forecasts and GDPS analyses in the 60\degree S
  and 90\degree S latitude bands at 70 hPa.  Weather forecast at
  3-5-10-day lead times from the experiments with (red) and without
  (blue) the inclusion of ozone radiative coupling.  The Y-axis shows
  differences in degrees (left panel) and m/s (right panel).  Marks on
  the right axis denote the mean differences (\%) and the number in
  brackets denotes the percentage of forecasts for which the
  differences are larger for the uncoupled experiment.  The dotted
  lines are the linear regression that best fits the data of both
  experiments.}
\label{ts_tu_70hpa_sp}
\end{figure}

The impact of ozone heating on medium-range weather forecasts in the
polar regions can be evaluated by comparing an ozone-coupled
experiment with a control cycle in which the ozone monthly mean
climatologies are used for the computation of the heating
rates. Figure~\ref{ts_tu_70hpa_sp} (left panels) shows the impact of
ozone heating on the mean temperature forecasts at 70 hPa over the
Antarctic region at different lead times.

\begin{figure}
\includegraphics[width=16cm]{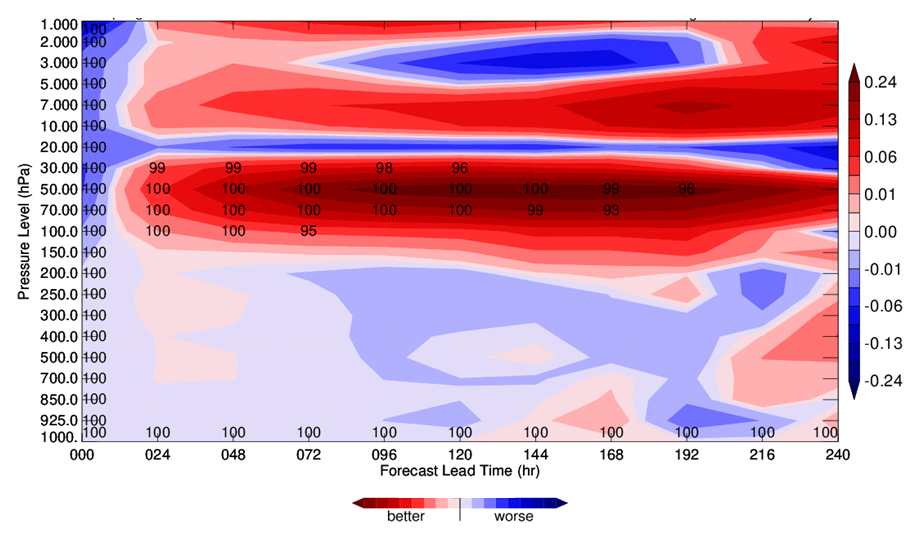}
\caption{Temperature standard deviation differences between forecasts
  and GDPS analyses as a function of lead time in the 60\degree S and
  90\degree S latitude bands for the October-December 2020 period for
  the experiments with and without ozone coupling.  Better results
  using the ozone coupled model (red) and better results using
  climatological ozone (blue). Numbers represent the confidence level
  computed using permutation tests (see chapter 15 of
  \citet{Efron1994}).}
\label{std_tt_sp}
\end{figure}

The timeseries indicate that the radiative impact of ozone starts to
become significant around mid-September and increases steadily until
the final breakup of the polar vortex. Prior to that period, solar
heating rates are relatively small, which mainly explains the weak
impact of ozone on temperature forecasts. In November and December,
the solar fluxes reach their maximum intensity while the polar vortex
becomes highly perturbed. Results show that throughout this period,
the differences increase with forecast lead times by about 0.1 K/day
with the use of prognostic ozone. This contributes to eliminating most
of the warm temperature bias present in the control experiment at all
lead times.

\begin{figure}
\includegraphics[width=8cm]{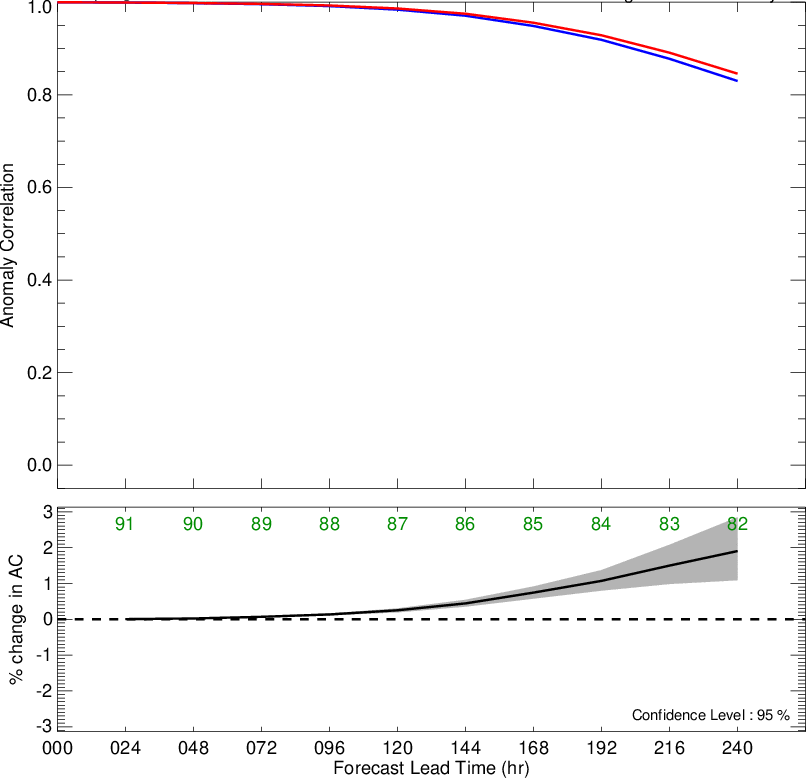}
\caption{Temperature anomaly correlation at 70 hPa within the
  60\degree S and 90\degree S latitude bands for the period
  October-December 2020 as a function of lead times from the
  experiments with (red) and without (blue) the inclusion of ozone
  radiative coupling.  Digits in green represent the number of
  forecasts used in the calculation at the given lead time.  The
  bottom panel shows the differences between both experiments whereas
  the shading indicates the 95\% confidence level.  Confidence limits
  are obtained from bootstrapping the adjusted correlation coefficient
  after applying a Fisher transformation to convert the correlation
  coefficient to a normally distributed variable. This process
  involves repeatedly resampling the data with replacement.}
\label{acc_tt_70hpa_sp}
\end{figure}

\begin{figure}
\includegraphics[width=16cm]{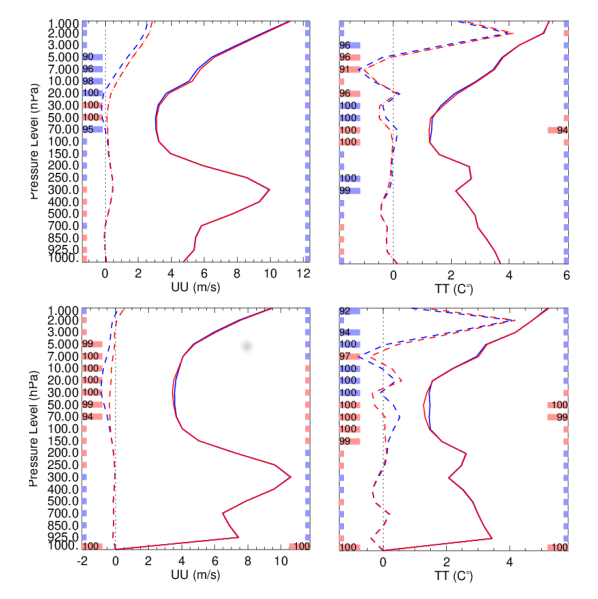}
\caption{Five-day weather forecasts against GDPS reanalyses in the
  60\degree N and 90\degree N latitude bands (top panels) and
  60\degree S and 90\degree S latitude bands (bottom panels) for the
  zonal wind (left) and temperature (right) from the experiments with
  (red) and without (blue) the inclusion of ozone radiative coupling.
  Dashed lines are mean biases while solid lines are error standard
  deviation.  Boxes on the left (right) denote statistical
  significance levels for the bias (standard deviation) computed using
  permutation tests (see chapter 15 of \citet{Efron1994}).  Red (blue)
  boxes mean that the ozone coupled (uncoupled) experiment is better.}
\label{pro_weather_5day}
\end{figure}

Figure~\ref{ts_tu_70hpa_sp} (right panels) reveals that the associated
cooling intensifies the strength of the polar vortex through thermal
wind balance. It contributes to reducing the zonal wind bias, which is
particularly significant throughout the vortex closure period in
December. The impact on the mean temperature and mean zonal wind fields
is statistically significant and  represent a significant fraction of the
standard deviation as shown in Figure~\ref{pro_weather_5day}.

The timeseries in Figure~\ref{ts_tu_70hpa_sp} also indicate that the
impact of the heterogeneous chemistry parameterization scheme on the
quality of weather forecasts at medium-range timescales is relatively
small. Results indeed show that the temperature signal is minimal in
August and September when the parameterization scheme is effective,
whereas the temperature signal increases significantly from early
October once the ozone hole is formed. These results suggest that the
improvement in the temperature and wind forecasts is linked to the
representation of the ozone advection process and the quality of the
ozone analyses, which mainly determined the quality of the ozone
forecast and associated radiative forcing during the ozone recovery
period from early October.

\begin{figure}
\includegraphics[width=16cm]{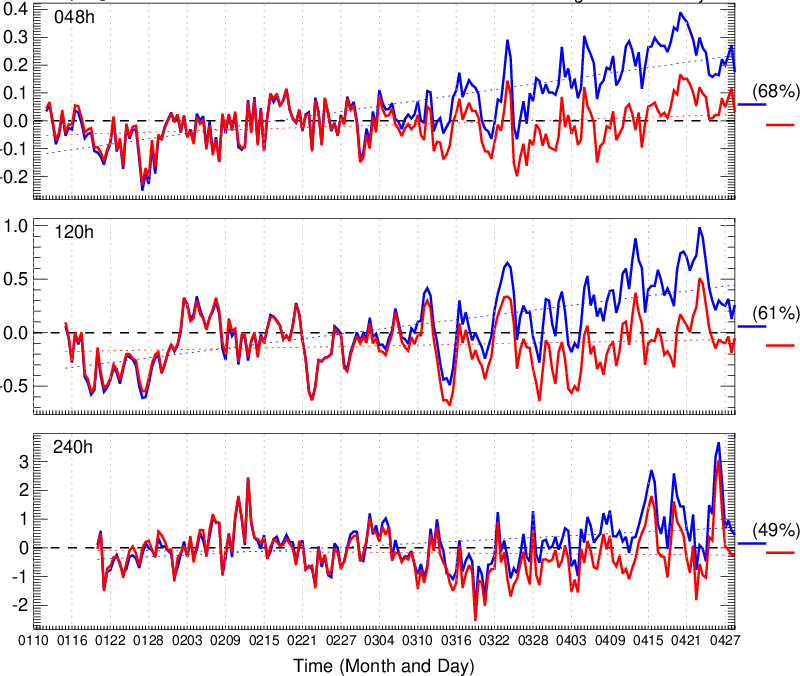}
\caption{Time Series of temperature mean differences (degrees) between
  forecasts and GDPS analyses in the 60\degree N and 90\degree N at
  70 hPa.  Weather forecast at 3-5-10-day lead times from the
  experiments with (red) and without (blue) the inclusion of ozone
  radiative coupling.  Marks on the right axis denote the mean
  differences (\%) and the number in brackets denote the percentage of
  forecasts for which the differences are larger for the uncoupled
  experiment.  The dotted lines are the linear regression that best
  fits the data of both experiments.}
\label{ts_tt_70hpa_np}
\end{figure}

During the ozone recovery period from early October, the temperature
standard deviation between the forecast and the analyses decreases
systematically throughout the stratosphere with the use of the ozone
coupled model. Figure~\ref{std_tt_sp} compares the mean differences
between the temperature standard deviation from both experiments at
all lead times throughout the model domain. It shows that the largest
impact on the temperature forecasts occurs in the lower stratosphere,
where the ozone distribution is far from zonal mean conditions.
Results indicate that the reduction of the mean temperature standard
deviation exceeds 0.2 degrees at 50 hPa and maximizes between 4-day
and 8-day lead times.

The temperature response shown in the previous figures can be compared
with published results from \citet{Monge2022}. This study shows annual
mean temperature differences that can reach 1\degree at a 10-day lead
time in the lower stratosphere extra-tropical regions, which is
comparable to the signal seen in Figure~\ref{ts_tu_70hpa_sp}.  Their
study also indicates that the temperature signal in the Southern
Hemisphere maximizes around the 5-day lead time, similar to the signal
shown in Figure \ref{std_tt_sp}. This may suggest some radiative
adjustment of the temperature beyond that timescale in the region.

\begin{figure}
\includegraphics[width=16cm]{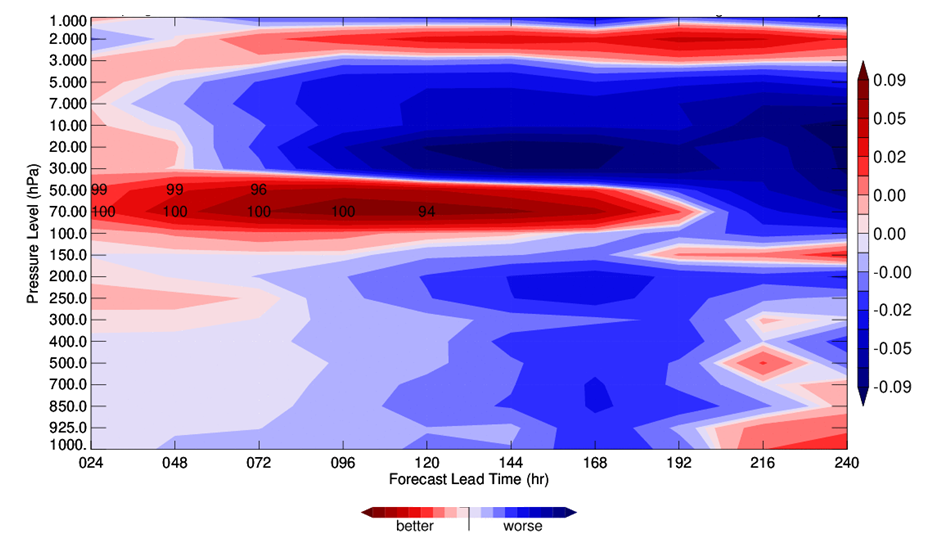}
\caption{Temperature standard deviation differences between forecasts
  and GDPS analyses as a function of lead time in the 60\degree N and
  90\degree N latitude bands from January 10$^{th}$ to April
  29$^{th}$ 2020 for the experiments with and without ozone coupling.
  Better results using the ozone coupled model (red) and better
  results using climatological ozone (blue). Numbers represent the
  confidence level computed using permutation tests (see chapter 15 of
  \citet{Efron1994}).}
\label{std_tt_np}
\end{figure}

Figure \ref{acc_tt_70hpa_sp} shows that the temperature anomaly
correlation at 70 hPa over the Antarctic region increases
significantly at all forecast lead times during this period. The
impact is particularly notable beyond the 3-day lead time, where the
signal exceeds the 95\% confidence level. This result shows that
temperature predictability is relatively high in the region and
increases by almost half a day at the 10-day lead time with the use of
the ozone-coupled model, indicating a significant improvement in
forecast errors throughout the event.

Over the Arctic, the radiative impact of ozone is generally smaller
but remains notable. The temperature signal maximizes during
March-April when solar heating becomes significant over the region.
Figure~\ref{ts_tt_70hpa_np} shows that the model suffers from a warm
bias over the Arctic, which is reduced with the use of a prognostic
ozone representation. The signal remains significant until the end of
April, which is due to the very unusual persistence of ozone-depleted
air masses within the Arctic vortex throughout the period
\citep{Manney2020}.

Figure \ref{std_tt_np} also shows a decrease in the mean temperature
standard deviation throughout the period. The reduction is
statistically significant throughout the lower stratosphere and
reaches 0.1 degrees around 70 hPa. The radiative impact of ozone
decreases faster than in the Southern Hemisphere and doesn’t last
beyond the 8-day lead time. It is noted that the increase in the
standard deviation in the upper stratosphere is not statistically
significant at the 95\% confidence level. 

The overall benefit of using an ozone coupled model for weather
forecasting in both polar regions is shown in
Figure~\ref{pro_weather_5day}. The results indicate that the
temperature forecasts at a 5-day lead time are significantly colder
throughout the lower stratosphere in both polar regions, stemming from
a better representation of the ozone loss process. In the SH, the
weather forecasts are generally improved with a more intense and
colder polar vortex in better agreement with the analyses. The
standard deviation is also improved throughout the lower stratosphere,
and the signal exceeds the 95\% confidence level in the case of
temperature. In the NH, the model response is similar, with a colder
polar vortex and stronger zonal wind throughout the springtime
period. In this case, the departure between the analyses and the
forecasts has slightly increased throughout most of the stratosphere,
which can be associated with the strengthening of the Brewer-Dobson
Circulation. This shows the model’s limitation in comprehensively
resolving the nature of the momentum forcings and interactions
involved.


\section{Conclusions} \label{sec:conclusions}

The use of ozone as a prognostic constituent can improve NWP systems
in several ways. It is a well-observed constituent that can be used as
a valuable metric for evaluating the quality of medium-range weather
forecasts throughout the stratosphere. The prognostic ozone
distribution can also be used for the computation of heating rates,
which further contributes to improving the predictability of
large-scale weather variables. The impact of including ozone radiative
coupling is particularly important in polar regions where the zonal
asymmetries of the ozone distribution represent a significant element
of the radiative forcing.

In this study, we use the ozone coupled ECCC GDPS to assess the impact
of ozone radiative forcing on the accuracy of ozone and weather
forecasts during major springtime ozone depletion events in both polar
regions in 2020. The NWP model employs a linearized gas-phase
photochemical module along with a simplified parameterization scheme
to represent the ozone loss associated with the important
heterogeneous reactions occurring during polar ozone depletion events.

The study shows that the predictability of TOC is relatively high
during ozone depletion events, exceeding 10 days in both polar
regions. In the SH, results indicate that ozone mean biases against
GDPS analyses decreased by about 75\% in August and September
throughout the lower stratosphere with the use of the heterogeneous
chemistry parameterization scheme. Implementing this process in the
model has also contributed to improving TOC predictability by about 8
hours during the same period, indicating a significant reduction in
ozone modeling errors. In the NH, the parameterization scheme has
helped reduce TOC biases by about 40\% in February and March at all
lead times, including a few percent reduction in the standard
deviation.

The use of ozone as a radiatively active constituent had a significant
impact on the quality of weather forecasts throughout the lower
stratosphere in both polar regions. In the SH, the use of an
ozone coupled NWP model has contributed to improving the temperature
and zonal wind distribution in the forecasts at all lead times. The
study shows that the mean temperature forecast during the SH
springtime period has decreased by about 0.5 degrees at 5-day lead
times, which has also contributed to increasing the zonal wind field
by about 0.5 m/s through thermal wind balance, in better agreement
with GDPS analyses. The inclusion of ozone radiative coupling has also
contributed to increasing temperature predictability by about 12 hours
at 70 hPa, demonstrating the model’s capability to resolve the
radiative and photochemical interactions between temperature and ozone
at those timescales.  Over the Arctic the radiative impact of ozone is
significantly weaker due to the large dynamical variability of the
region. During the 2020 winter period, the impact of ozone radiative
coupling on weather forecasts is particularly significant in March,
when ozone depletion processes and ozone heating become
significant. In such conditions, the use of an ozone coupled
forecasting system reduces the temperature bias that develops at short
lead times with the use of a prescribed ozone climatology. The results
also show that most of the improvement in temperature and wind
forecasts within 10-day lead times in both hemispheres is associated
with the quality of ozone analyses, which serve as initial conditions,
and the representation of ozone transport processes in the NWP model.

The study highlights the importance of improving the representation of
ozone heating within NWP forecasting systems. It shows that the
radiative impact of ozone over polar regions develops rapidly and can
significantly improve the quality of weather forecasts. The study
demonstrates that the coupling mechanisms between ozone and
temperature can be represented at medium-range timescales with the use
of ozone assimilation and a simplified modeling approach.

The results also indicate that the radiative impact of ozone on
weather increases steadily over time and will likely affect weather
predictability at monthly and seasonal timescales. At those
timescales, the influence of the ozone initial conditions will
decrease, and the quality of the ozone forecasts will be mainly
determined by modeling errors. In this context, the use of
ozone coupled systems for monthly and seasonal forecasting will likely
necessitate the development of more comprehensive modeling approaches
(e.g. \citet{Monge2011}) that will further extend ozone
predictability throughout the UTLS region.

\appendix

\section{On the representation of ozone over the Arctic} 

The ozone hole area, defined as the region encircled by the TOC 220 DU
contour, is typically used for characterizing the extent of the ozone
depletion process. The size of the area varies significantly and
reached a maximum of slightly over 1 million km$^2$ in mid-March
\citep{Dameris2021}. The GDPS analysis on March 28$^{th}$ (Fig.A1 -
left panel) shows that the ozone hole is located north of Greenland,
in good agreement with OMI measurements (Fig.A1 - right panel), even
though the TOC values appear slightly overestimated by the
GDPS. Fig.A2 shows the GDPS ozone profile for the same day at
Ny-Alesund, which is compared against an in-situ ozonesonde
observation. Results show that the signature of the ozone depletion
around 70 hPa is well represented in the analyses.

\setcounter{figure}{0} \renewcommand{\thefigure}{A \arabic{figure}}

\begin{figure}
\includegraphics[width=16cm]{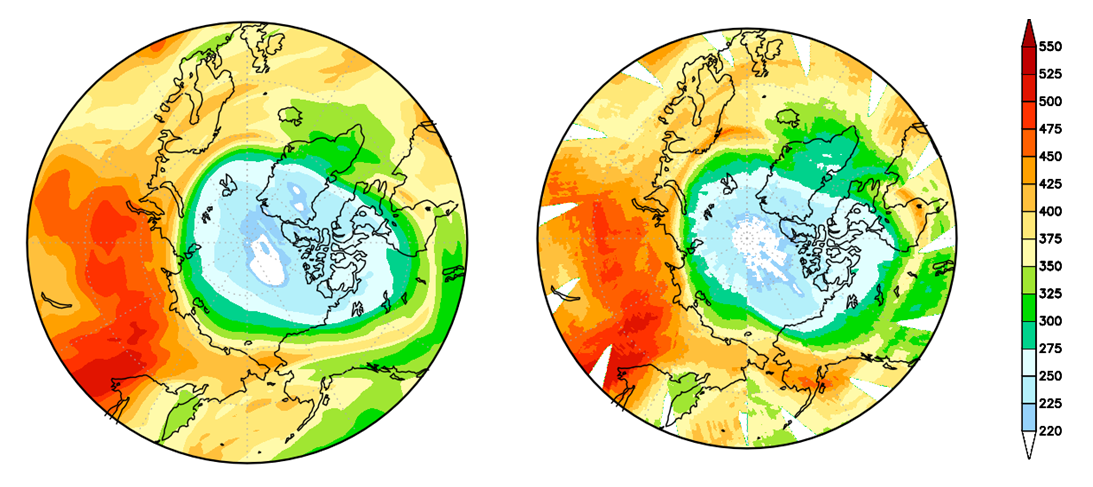}
\caption*{Fig.A1. TOC (DU) over the Arctic on March 28$^{th}$ 2020 from (left) GDPS analyses and (right) OMI .}
\end{figure}

\begin{figure}
\includegraphics[width=8cm]{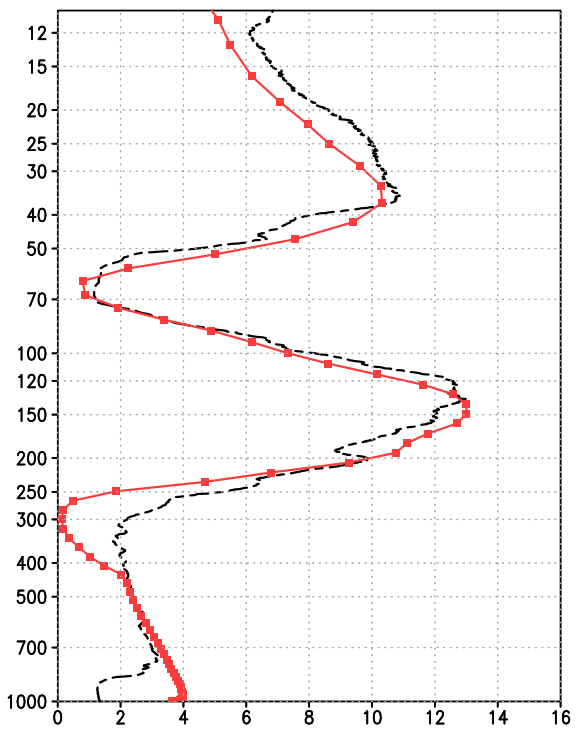}
\caption*{Fig.A2. Ozone profiles at Ny Alesund (78.9N, 11.9E) on March 28$^{th}$ 2020 from the GDPS (red) and the ozonesonde (black).}
\end{figure}

\clearpage


\bibliographystyle{ametsocV6}
\bibliography{references.bib}

\clearpage

\setcounter{figure}{0} \renewcommand{\thefigure}{\arabic{figure}}

\end{document}